\documentclass[a4paper, 11pt, oneside]{article} 

\usepackage{geometry}
\usepackage{booktabs} 
\usepackage{breakurl}
\usepackage{mathptmx} 

\usepackage{fancyhdr}
\usepackage[normalem]{ulem}
\usepackage{microtype}
\usepackage{graphicx}
\usepackage{subfigure}
\usepackage{colortbl}
\definecolor{mygray}{gray}{.88}
\usepackage{multirow}
\usepackage{amsmath}
\usepackage{bm}
\usepackage{epsfig}
\usepackage{authblk}

\usepackage[hyphens]{url}
\usepackage{hyperref}
\usepackage{color}
\usepackage{soul}
\usepackage{bbding}
\definecolor{mygray}{gray}{.88}

\begin{document}

\begin{titlepage} 

	\centering 
	
	\scshape 
	
	\vspace*{\baselineskip} 
	
	
	\rule{\textwidth}{1.6pt}\vspace*{-\baselineskip}\vspace*{2pt} 
	\rule{\textwidth}{0.4pt} 
	
	\vspace{0.75\baselineskip} 
	
	{\LARGE BenchCouncil's View on Benchmarking AI and Other Emerging Workloads} 
	
	\vspace{0.75\baselineskip} 
	
	\rule{\textwidth}{0.4pt}\vspace*{-\baselineskip}\vspace{3.2pt} 
	\rule{\textwidth}{1.6pt} 
	
	\vspace{2\baselineskip} 
	
	
	
	\vspace*{3\baselineskip} 
	
	
	 \begin{flushleft}Edited By\end{flushleft}
	
	\vspace{0.5\baselineskip} 
	
	{ \begin{flushleft}Jianfeng Zhan \textit{(BenchCouncil Steering Committee Chair)} \\ Lei Wang  \textit{(BenchCouncil Big Data and CPU Tracks Executive Committee Co-chair)}\\ Wanling Gao \textit{(BenchCouncil Datacenter AI Track Executive Committee Co-chair)}\\Rui Ren \textit{(BenchCouncil Testbed Co-chairs)} \end{flushleft}}
	
	
	\vspace{0.5\baselineskip} 

	\vfill 
	
	
	\epsfig{file=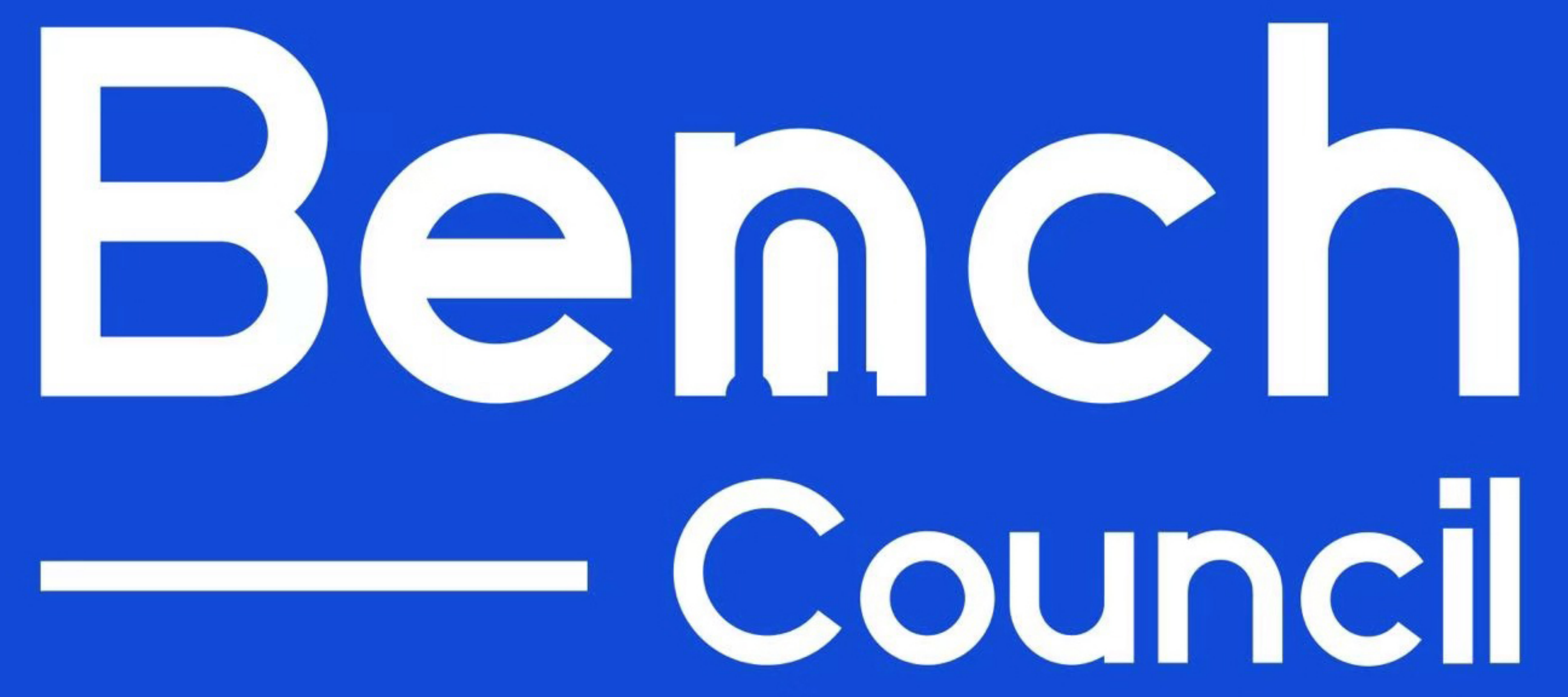,height=2cm}
	\textit{\\BenchCouncil: International Open Benchmark Council\\http://www.benchcouncil.org} 
	\vspace{5\baselineskip} 

	Technical Report No. BenchCouncil-BCView-2019 
	
	{\large Nov 12, 2019} 

\end{titlepage}


\title{BenchCouncil's View on Benchmarking AI and Other Emerging Workloads}

\author[1]{Jianfeng Zhan\thanks{Jianfeng Zhan is the corresponding author.}}
\author[1]{Lei Wang}
\author[1]{Wanling Gao}
\author[1]{Rui Ren}

\affil[1]{BenchCouncil (International Open Benchmark Council)}

\date{Nov 12, 2019}
\maketitle

\begin{abstract}
This paper outlines BenchCouncil's view on the challenges, rules, and vision of benchmarking modern workloads like Big Data, AI or machine learning, and Internet Services.  We conclude the challenges of benchmarking modern workloads as FIDSS (\underline{F}ragmented, \underline{I}solated, \underline{D}ynamic, \underline{S}ervice-based, and \underline{S}tochastic), and propose the PRDAERS benchmarking rules that the benchmarks should be specified in a \underline{p}aper-and-pencil manner, \underline{r}elevant, \underline{d}iverse, containing different levels of \underline{a}bstractions, specifying the \underline{e}valuation metrics and methodology, \underline{r}epeatable, and \underline{s}caleable. We believe proposing simple but elegant abstractions that help achieve both efficiency and general-purpose is the final target of benchmarking in future, which may be not pressing. In the light of this vision, we shortly discuss BenchCouncil's related projects.
\end{abstract}
\section{Challenges}
Our society is increasing relying upon information infrastructure, which consists of massive IoT, edge devices, extreme-scale datacenters, and high-performance computing systems. Those systems collaborate with each other to handle big data and leverage AI technique, and finally provide Internet services for huge end users with guaranteed quality of services. From a perspective of workload characterization, the emerging workloads like Big Data, AI, and Internet Services raise serious FIDSS (\underline{F}ragmented, \underline{I}solated, \underline{D}ynamic, \underline{S}ervice-based, and \underline{S}tochastic) challenges, which are significantly different from the traditional workloads characterized by SPECCPU (desktop workloads)~\cite{SPECCPU2017}, TPC-C~\cite{tpcc}, TPC-Web (Traditional web services)~\cite{tpc-w}, and HPL (high performance computing)~\cite{hpl} benchmarks.

The first challenge is \underline{f}ragmented. There are huge fragmented application scenarios, a marked departure from the past. However, there is a lack of simple but elegant abstractions that help achieve both efficiency and general-purpose. For example, for database, the relation algebra demonstrates its generalized ability, and any complex query can be written using five primitives like select, project, product, union, and difference~\cite{codd1970relational}. However, in a new era of big data, hundreds or even thousands ad-hoc solutions are proposed to handle different application scenarios, most of which are termed with NoSQL or NewSQL. For AI, the same observation holds true. There are tens or even hundreds of organizations who are developing AI training or inference chips to tackle their challenges in different application scenarios, respectively ~\cite{mlperf2019}. Though domain-specific software and hardware co-design is promising ~\cite{2018_turing}, the lack of simple but unified abstractions has two side effects. On one hand, it is challenging to amortize the cost of building an ad-hoc solution. On the other hand, single-purpose is a structure obstacle to resource sharing.  Proposing simple but elegant abstractions that help achieve both efficiency and general-purpose is our final target of workload modeling, benchmarking, or characterization in future, which may be not pressing.

The second challenge is \underline{i}solated. The real-world data sets and workloads or even AI models are treated as first-class confidential issues, and they are hidden with giant Internet service providers' datacenters, and isolated between academia and industry, or even among different providers~\cite{gao2019aibench}. This kind of isolation finally poses a huge obstacle for our communities towards developing an open and mature research field~\cite{gao2019aibench2}.

The third challenge is the \underline{d}ynamic complexity. The dynamic complexity of emerging workloads is in terms with that the common requirements are handled collaboratively by datacenters, edge, and IoT devices. It is also manifested by that different distributions of data sets, workloads, machine learning or AI models may substantially affect the system's behaviors.  Unfortunately, the system architectures are undergoing fast evolutions in terms of the interactions among IoT, edge, and datacenters.

The fourth challenge is the side effect of \underline{s}ervice-based architecture. On one hand, the software-as-a-service (SaaS) development and deploy model makes the workloads change very fast (so-called workload churn) ~\cite{Barroso2009The}, and it is not scalable or even impossible to create a new benchmark or proxy for every possible workload ~\cite{gao2018motif}. On the other hand, modern Internet services adopt a microservice-based architecture, which is often distributed across different datacenters, and consist of diversity of various modules with very long and complex execution paths. As the worst-case performance (tail latency) ~\cite{dean2013tail} does matter, the micro-service-based architecture also pose a serious challenge to benchmarking ~\cite{gao2019aibench}.

The final but not least challenge is the side effect of the \underline{s}tochastic nature of AI. The AI techniques are widely used to augment modern products or Internet services. The nature of AI is stochastic, allowing multiple different but equally valid solutions ~\cite{mattson2019mlperf}. The uncertainness of AI is manifested by adverse effect of lower-precision optimization on the quality of the final model, effect of scaling training on time-to-quality, and run-to-run variation in terms of epochs-to-quality~\cite{mattson2019mlperf}. However, the benchmarks mandates being repeatable, reliable, and reproducible. This conflict raises serious challenges.
\section{BenchCouncil Benchmarking Rules}
After revisiting the previous successful benchmarks, we propose the PRDAERS benchmarking rules as follows.

First, the common requirements should be specified only algorithmically in a \underline{p}aper-and pencil approach. This rule is firstly proposed by the NAS parallel benchmarks ~\cite{bailey2011parallel}, and well-practiced in the database community.  Interestingly, this rule is often overlooked by the architecture and system communities. Following this rule, the benchmark specification should be proposed firstly and reasonably divorced from individual implementations. In general, the benchmark specification should define a problem domain in a high-level language.

Second, the benchmark should be \underline{r}elevant~\cite{gray1992benchmark}. On one hand, the benchmark should be domain-specific~\cite{levine1997tpc} and distinguish between different contexts like IoT, edge, Datacenter and HPC.  Under each context, each benchmark should provide application scenarios abstracted from details, sematic-preserving data sets, or even quality targets (for AI tasks) that represent real-world deployments ~\cite{mlperf2019}. On the other hand, the benchmark should be simplified. It is not a copy of a real-world application. Instead, it is a distillation of the essential attributes of a workload ~\cite{levine1997tpc}. Generally, a real-world application is not portable across different systems and architectures.

Third, the \underline{d}iversity and representativeness of a widely accepted benchmark suite are of paramount importance. On one hand, this is a tradition witnessed by the past. For example, SPECCPU 2017 contains 43 benchmarks. The other examples include PARSEC3.0 (30), TPC-DS (99). On the other hand, modern workloads manifest much higher complexity. For example, Google's datacenter workloads show significant diversity in workload behavior with no single ``silver-bullet'' application to optimize for ~\cite{kanev2015profiling}. Modern AI models vary wildly, and a small accuracy change (e.g., a few percent) can drastically change the computational requirements (e.g., 5-10x) ~\cite{mlperf2019,bianco2018benchmark}. For modern deep learning workloads, someone may argue running an entire training session is costly, so few benchmarks should be included for reducing the cost. However, we believe the cost of execution time cannot justify including only a few benchmarks. Actually, the cost of execution time for other benchmarks (like HPC, SPECCPU on simulators) is also prohibitively costly.  So, for workload characterization, diverse workloads should be included to exhibit the range of behavior of the target applications, or else it will oversimplify the typical environment ~\cite{levine1997tpc}. On the other hand, for performance ranking (benchmarketing), it may be reasonable for us to choose a few representative benchmarks  for reducing the cost just like that the HPC Top500 ranking only reports HPL, HPCG, and Graph500 (three benchmarks out of 20+ representative HPC benchmarks like HPCC, NPB).

Fourth, the benchmarks should contain different levels of \underline{a}bstractions, and usually a combination of micro, component and end-to-end application benchmarks is preferred to. From an architectural perspective, porting a full-scale application to a new architecture at an earlier stage is difficult and even impossible~\cite{bailey1991parallel}, while using micro or component benchmarks alone are insufficient to discover the time breakdown of different modules and locate the bottleneck within a realistic application scenario at a later stage~\cite{bailey1991parallel}. Hence, a realistic benchmark suite should have the ability to run not only collectively as a whole end-to-end application to discover the time breakdown of different modules but also individually as a micro or component benchmark for fine tuning hot spot functions or kernels~\cite{gao2019aibench} .

Fifth, it should specify the \underline{e}valuation metrics and methodology. The performance number should be simple, linear, orthogonal, and monotonic~\cite{levine1997tpc}. Meanwhile, it should be domain relevant. For example, the time-to-quality (i.e., state-of-the-art accuracy) metric is relevant to the AI domain, because some optimizations immediately improve throughput while adversely affect the quality of the final model, which can only be observed by running an entire training session~\cite{mattson2019mlperf}.

Sixth, the benchmark should be \underline{r}epeatable, reliable, and reproducible~\cite{anglesLDBC}. Since many modern deep learning workloads are intrinsically approximate and stochastic, allowing multiple different but equally valid solutions~\cite{mattson2019mlperf}, it will raise serious challenges for AI benchmarking.

Finally, but not least, the benchmark should be \underline{s}caleable~\cite{gray1992benchmark}, and the benchmark users can scale up the problem size, so the benchmark is applicable for running on both small and large systems. However, this is not trivial for modern deep learning workloads, as accommodating system scale even requires changing hyperparameters, which can affect the amount of computation to reach a particular quality target~\cite{mattson2019mlperf}.
\section{Our Viewpoint}
Domain-specific software-hardware co-design is promising as a single-purpose solution will achieve high energy-efficiency than that of a general-purpose one. However, we believe the final target in future is to propose simple but elegant abstractions that help achieve both efficiency and general-purpose for big data, AI, and Internet services.

To meet with the fragmented application scenarios, currently, BenchCouncil sets up (datacenter) Big Data, datacenter AI, HPC AI, AIoT, and Edge AI benchmarking tracks, and released BigDataBench~\cite{gao2018bigdatabench} (\url{http://www.benchcouncil.org/BigDataBench/index.html}) and AIBench ~\cite{gao2019aibench,gao2019aibench2}(\url{http://www.benchcouncil.org/AIBench/index.html}) benchmark suites for datacenter big data and AI, Edge AIbench ~\cite{hao2018edge}(\url{http://www.benchcouncil.org/EdgeAIBench/index.html}) for edge AI, AIoT Bench ~\cite{luo2018iot}(\url{http://www.benchcouncil.org/AIoTBench/index.html})for IoT AI, and HPC AI500 ~\cite{jiang2018hpc} (\url{http://www.benchcouncil.org/HPCAI500/index.html}) for HPC AI. Those benchmarks are in fast evolution. We release the source code, pre-trained model, and container-based deployment on the BenchCouncil web site (\url{http://benchcouncil.org/testbed/index.html}).

On the other side, we propose an innovative approach to modeling and characterizing the emerging workloads. We consider each big data, AI and Internet service workload as a pipeline of one or more classes of unit of computation performed on different initial or intermediate data inputs, which we call a data motif~\cite{gao2018motif}. After thoroughly analyzing a majority of workloads in five typical big data application domains (search engine, social network, e-commerce, multimedia and bio-informatics), we identify eight data motifs that take up most of run time, including Matrix, Sampling, Logic, Transform, Set, Graph, Sort and Statistic ~\cite{gao2018motif}. We found the combinations of one or more data motifs with different weights in terms of runtime can describe most of big data and AI workloads we investigated ~\cite{gao2018motif,gao2018proxy}. In conclusion, the data motifs are promising as simple but elegant abstractions that achieve both efficiency and general-purpose.

On the basis of the data motif methodology, we are proposing a new benchmark suite, named BENCHCPU ~\cite{benchcpu}, to characterize emerging workloads, including Big Data, AI, and Internet Services. The goal of BENCHCPU is to abstract ISA (Instruction Set Architecture) independent workload Characterizations from emerging workloads. BENCHCPU will be portable across edge, IoT, and datacenter processor architectures. Furthermore, we are working on an open-source chip project, named EChip. On the basis of data dwarf approaches, the goal of EChip is to design an open source general-purpose ISA for emerging workloads, which is a marked departure from single-purpose accelerators. The ISA of EChip is composed of the general-purpose instruction set and the domain-specific instruction set. The general-purpose instruction set is modular-based basic instruction set, which always remains unchanged (Minimum implementation subset) and is compatibility with Linux ecosystems. The domain-specific instruction set is an extension of domain-specific customization, proposed for emerging workloads like big data, AI, internet service.
\section{Revisiting the Benchmarking Principles and Methodology}
Benchmarking principle defines the rules and guidelines on what the important criteria need to be considered for a good benchmark~\cite{anglesLDBC,zhao2015revisiting}. The benchmarking methodology specifies systematic strategies and processes on how to construct a benchmark. Jim Gray summarizes four key criteria to define a good benchmark---relevant, portable, scalable, and simple~\cite{gray1993database}. Also, other benchmark consortiums propose their own principles and methodologies.

The TPC Benchmarks are a series of domain-specific benchmarks targeting measuring transaction processing (TP) and database (DB) performance. They believe that a domain-specific benchmark should satisfy three criteria~\cite{levine1997tpc,huppler2009art}: (1) No single metric can measure the application performance of all domains, (2) The more general the benchmark, the less useful it is for anything in particular, (3) A benchmark is a distillation of the essential attributes of a workload. They adopt a benchmark methodology with the concept of "functions of abstraction" and "functional workload model", which abstract the compute units that frequently appeared with repetitions or similarities.

SPEC benchmarks are a set of benchmarks for the newest generation of computing systems~\cite{spec}, guided by six implicit design principles: application-oriented; portable; repeatable and reliable of benchmarking results; consistent and fair across different users or different systems; diverse workloads that can run independently, and reporting unit of measurement, e.g., throughput.

PARSEC (Princeton Application Repository for Shared-Memory Computers) is a benchmark suite for chip multiprocessors. The PARSEC benchmarks are constructed following five requirements~\cite{bienia2011benchmarking}. (1) The benchmarks should be multi-threaded. (2) Emerging workloads should be considered. (3) The benchmarks should cover diverse applications and a variety of platforms, and accommodate different usage models. (4) The benchmarks should use state-of-the-art algorithms and data structures. (5) The benchmarks should be designed to support research.

\section{Conclusion}
This paper outlines BenchCouncil's view on the challenges, rules, and vision of benchmarking modern workloads. We conclude the challenges of benchmarking modern workloads as \underline{F}ragmented, \underline{I}solated, \underline{D}ynamic, \underline{S}ervice-based, and \underline{S}tochastic. After revisiting previous successful benchmarks, we propose the PRDAERS benchmarking rules that the benchmarks should be specified in a \underline{p}aper-and-pencil manner, \underline{r}elevant, \underline{d}iverse, containing different levels of \underline{a}bstractions, specifying the \underline{e}valuation metrics and methodology, \underline{r}epeatable, and \underline{s}caleable. We believe proposing simple but elegant abstractions that help achieve both efficiency and general-purpose is the final target of benchmarking in future, which may be not pressing.  In the light of this vision, we shortly discuss BenchCouncil's related projects, including BigDataBench, AIBench, HPC AI500, Edge AIBench, AIoT Bench, and BENCHCPU, and EChip projects.



\end{document}